\pdfoutput=1
\documentclass[12pt]{article}

\usepackage{graphicx,amsmath,amssymb,pslatex,natbib}

\newcommand{\be}[1]{\begin{equation}\label{#1}}
\newcommand{\ee}{\end{equation}}

\begin{document}
\begin{center}{\Large How sensitive are equilibrium pricing models to 
real-world distortions?} \end{center}

\begin{center}

\underline{HARBIR LAMBA}\footnote{Department of Mathematical Sciences,
George Mason University, MS 3F2, 4400 University Drive,
Fairfax, VA 22030 USA}\\

\end{center}

\begin{abstract}

In both finance and economics, quantitative models are usually studied as
isolated mathematical objects --- most often defined by very strong
simplifying assumptions concerning rationality, efficiency and the
existence of disequilibrium adjustment mechanisms. This
raises the important question of how sensitive such models might be to real-world
effects that violate the assumptions.

We show how the consequences of rational behavior caused by perverse
incentives, as well as
various irrational tendencies identified 
 by behavioral economists, can be
systematically and consistently introduced into an agent-based model
for a financial asset.  
This generates a class of models which, in the special case where such
effects are absent, reduces to
geometric Brownian motion --- the usual equilibrium
pricing model. Thus we
are able to numerically perturb a widely-used equilibrium pricing
model  market and investigate
its stability. 

The magnitude of such perturbations in real markets can be
estimated and the simulations imply that this is far outside the
stability region of the equilibrium solution, which is no longer
observed. Indeed the price fluctuations generated by endogenous
dynamics, are in good general agreement
with the excess kurtosis and heteroskedasticity of actual asset
prices.

The methodology is presented within the context of a financial
market. However, there are close links to concepts and theories
from both micro- and macro-economics including rational expectations,
Soros' theory of reflexivity, and Minsky's theory of financial
instability. 

% Finally we describe formal links
% between the endogenous cascade processes generated by the model (that are
% responsible for the largest price-changes occurring in the system)
% with the physics
% of earthquakes and the mathematics of Queueing Theory. This may have
% implications for the modelling of rapid-deleveraging in financial markets.

\end{abstract}



\section{\label{intro}Introduction}  

All of the mainstream schools of economic thought and, in turn, modern finance have been
strongly motivated by the mathematical 
elegance and predictive success of Newtonian mechanics 
and statistical physics. This has resulted in 
mathematical models constructed so that the solution is a unique, stable,
equilibrium whose value is both history-independent and a continuous
function of the system variables.

The notion of equilibrium is close to being a unifying concept
in modern economics (see \cite{s97a,n72}) although its precise meaning is
contextual.  Here we shall take it to mean
the absence (or complete cancellation) of endogenous dynamics so that
the state of the system is 
determined solely by the current values of exogenous variables and
not, say, on any prior state of the system. This is certainly the case
for gas molecules in a closed box that are reacting to slow changes in
temperature --- the pressure is determined by the Ideal
Gas Law and the current temperature. This happy state of
affairs is guaranteed by  the Laws of Thermodynamics
%(Conservation of Energy and Increasing Entropy respectively) 
that
provide  both a unique statistical equilibrium and a physical
mechanism for reaching it. 

One may speculate on what would happen if there was a special type of
molecule that did not obey the First and
Second Laws of Thermodynamics (Conservation of Energy and Increasing
Entropy respectively). These particles can spontaneously
speed up or slow down, reveal a temporary preference for one
side of the box over the other, or imitate the particles close to them. 
One may then wonder which of these two types of molecule
has more in common with agents in an economic system.  

Economic systems are aggregations of many heterogeneous
agents  and the a priori requirement of
history-independent equilibrium 
solutions has profound consequences (see \cite{Kirman}). Most models are
generated by assuming  that agents display enough homogeneity to be 
 `averaged' or scaled-up and so replaced by a representative
agent who is memoryless  and is both perfectly rational (usually in
the sense of maximizing some hypothetical utility function) and correct about
future expectations. This averaging procedure, when applied to
expectations, is known as the Rational Expectations Hypothesis  and
implies that
while individual agents' expectations may be wrong
they all use all available information and, on average, agree with the
expectations assumption being used 
with no systematic deviations. 
\footnote{Such averaging assumptions are the economic analogue of the Central
Limit Theorem which states that the average of independent random
variables (from a distribution with finite mean and variance) is
normally distributed. The assumption of independence is crucial!}

In this paper we shall consider a heterogeneous agent model of
a financial market to demonstrate that there are significant, systemic, real-world
effects that cannot be removed by averaging once they reach a certain
(rather low) level. This is the point at which positive feedback
mechanisms becomes strong enough to defeat the equilibrating
process. Then the equilibrium 
market solution loses stability completely and complex internal
(endogenous) 
dynamics develop. This alters the future evolution of the
system in a history-dependent way that cannot be adequately represented by any
equilibrium model --- although for much of that time the system may
easily be mistaken for one that is in equilibrium with an underlying trend.
Finally, the endogenous dynamics rapidly reverse in a cascade process.

The main source of positive feedback/instability (and
non-averagability) 
in the system is {\em herding
  behavior} whereby agents  have some additional motivation (ranging from completely
irrational to hyper-rational) to prefer the position taken by the
majority of agents. While there are many possible causes of herding
behavior, the effects should usually be similar and can be modelled
very simply using the framework developed in Section~\ref{model}. 

Few would argue with the statement that irrational behavior and
herding are a common feature of the biggest market bubbles and
crashes. Yet the question remains as to how such factors may affect
the workings of a market when, even with the benefit of hindsight, no
obvious mispricings are occurring. 
Two previous studies (\cite{ay85,ss90})  looked at the effects
of small changes to otherwise maximizing rational behavior. They both
showed that even small changes can cause significant (first order)
changes to the value of the equilibrium. The second study is
particularly relevant in that it considered rational herding by
investment managers concerned about their relative performance 
to be  the primary source of the deviation. However
these analyses were performed within an equilibrium framework and so
precluded the possibility of non-equilibrium dynamic
solutions. Similarly \cite{b92} and \cite{bhw92} introduced models of herd behavior
via sequential decision-making but again within an equilibrium framework.

The modelling framework  to be described in Section~\ref{model} was introduced in
\cite{ls08} and \cite{l10} and is based upon earlier related models
(\cite{cgls05,cgl06,cgls07,ls08}). 
The primary
motivation for this earlier work was to show how to incorporate various systemic
defects such as perverse incentives and investor psychology into an
otherwise efficient market and to establish a causal link
between agent behavior at the micro level and the non-Gaussian price
statistics observed in  financial markets. One clear conclusion   was that
herding can indeed induce `fat-tails'  consistent with observed
power-law decays of real asset price changes. Thus herding is a
possible, if not likely, contributor to an extremely important
phenomenon  that is 
inconsistent with standard financial models.

The motivation here is slightly different, although the modelling
details are similar. The model below should be
considered  a `stress-test' of  an extant equilibrium 
pricing model that is carried out by weakening a particular subset of
the assumptions. It is important to note that this allows us to
carry over, without any detailed justification, those assumptions from the standard
pricing 
model that are not being weakened.

The paper is organized as follows. The modelling framework is described fully in
Section~\ref{model} together with an explanation of how
various non-standard motivations, such as perverse incentives and the findings of
behavioral economics, 
can be approximated.  Then in Section~\ref{num1} the results of numerical simulations
are provided for realistic estimates of the model parameters. These
demonstrate that the 
qualitative and quantitative changes introduced by such
`imperfections' are consistent with observations of real markets. In
Section~\ref{links} various links are established between the
modelling philosophy with  concepts and theories from economics,
finance, 
mathematics and physics. Section~\ref{num2} contains the most
significant numerical results. Here the herding strength is used as
a bifurcation parameter to establish how much herding is required to
destabilize the equilibrium market solution. The answer would appear to
be far lower than a plausible
estimate of herding in financial
markets. Conclusions and suggestions for further research are given in
Section~\ref{conc}.

\section{\label{model}The Model}

The simplest, and most common, asset pricing model assumes that the
price $p(t)$ at time $t$ obeys the stochastic differential equation
(SDE)
\begin{equation} dp = a p \;dt + b p \;dB  \label{sde} \end{equation}
where $a$ and $b$ are the constant drift and volatility of the price per unit time and
$B(t)$ is a standard Brownian motion
representing the arrival of uncorrelated
Gaussian-distributed information. Equation  \eqref{sde} is justified
by positing that new information is instantaneously and perfectly
translated into a 
price change via some equilibrating process.
The solution to
\eqref{sde}, found using the It\^{o} Calculus, is the geometric Brownian motion
\begin{equation} p(t) =  p(0) \exp\left( (a-\frac12 b^2)t +
    b B(t)\right) \label{sdesol}. \end{equation}

The SDE \eqref{sde} is highly unusual in that it has an explicit
solution. Even more unusually, $p(t)$ depends only upon the
{\em current} value of $B(t)$ and not the entire history of the Brownian
process up to that point. Thus the variable $p(t)$ can be considered a paradigm for
other variables in economic models
that are assumed to be the  outcome of instantaneous equilibrating
processes that are history-independent (i.e. that have no memory).      

Without loss of generality, we may choose $a = 0$ so that $p(t)$
becomes  the price relative to the risk-free interest rate plus risk premium. 
We may also, by rescaling time, choose $b = 1$ and then 
discretize time so that the solution at the end of the $n^{\rm th}$
time interval of length $h$ is
  \begin{equation}
p(n)=p(n-1) \exp\left(\sqrt{h}\eta(n)
 -h/2\right)\label{price2}
\end{equation}
where $\eta(n) \sim {\cal N}(0,1).$

Note that the log-price $P(n) = \ln p(n)$ follows a standard Brownian
motion with the log-price changes $\Delta P_n = P(n)-P(n-1)$ having a
Gaussian distribution ${\cal N}(0,h)$.  However this is in very poor
agreement with reality. The `stylized facts' of financial markets (see
\cite{ms00,c01}) 
are a set of statistical observations that appear to hold across all
asset classes, independent of geography, history and trading
rules. There are  two highly significant deviations from
\eqref{price2}. The first is the presence of `fat-tailed' price
returns whereby the occurrence of the largest price changes (as
measured over
intervals of hours up to months or years) follows an approximate
power-law decay by contrast with the exponential decay of Normal
distributions. Thus the probability of the largest price moves is
underestimated by many orders of magnitude. The second phenomena is
volatility clustering, also known as heteroskedasticity, whereby
large price moves (in either direction)
are more likely to occur shortly after other large price moves. It is quantified by
calculating the autocorrelation of the volatility $|P(n)-P(n-1)|$ which
is again observed to follow a power-law decay. Under the Normal
approximation, this autocorrelation should be precisely zero.

Almost by definition, the price changes caused by the information
stream are effected by agents in the marketplace who a) act very fast
and b) are motivated by the arrival of new information $\eta(n)$. 
The standard neoclassical argument  is to
suppose that it is {\em as if} all agents are continuously, instantaneously and
correctly (on average) maximizing their respective utility functions.
In reality the presence of transaction costs and the immense
computational effort will mean that trading occurs over much longer
timescales, at least for a subset of $M$ agents. We shall call them
`slow agents'. We do not assume that slow
 agents are of uniform size (in terms of their trading positions) and
 thus weight the $i^{\rm th}$ slow agent by her size $w_i$ and define $W =
 \sum_{i=1}^M w_i$. 

Note that, in the standard pricing model \eqref{price2},
 the 
market-clearing  mechanism is assumed to be efficient and thus the
details are unimportant and not
specified. 
Similarly, only the slow agents will be explicitly simulated and it is
assumed that the fast agents provide sufficient liquidity\footnote{The
  reader is directed to \cite{l10} for a discussion of how the model can be modified at
  times of severe market stress when liquidity cannot be assumed.}.

We now make some assumptions that, it must be emphasized, 
 are not fundamental to the modelling philosophy. They keep the
 model simple and  are sufficient for the purpose at hand.
Firstly, we assume that over the $n^{\rm th}$ time interval the $i^{\rm th}$ slow
agents can only be in one of two 
states, the state $s_i(n)=+1$ meaning that she owns  $w_i$ units of the
asset, and the state $s_i(n)=0$ meaning that she owns none of the
asset\footnote{In reality a slow agent may choose to gradually
  increase or decrease their holdings, short the market, or buy
  derivatives, but this complicates
  the dynamics of the slow agents without providing further insights.}.
We can thus define the quantity $\sigma(n) = \frac{2}{W} \sum_{i=1}^M
s_i(n) w_i -1$ which is a linear measure of the aggregate demand of the slow
agents. Note that $\sigma(n) = -1$ when none of the slow agents own the asset
and  $\sigma(n) = +1$ when  they all do. Changes in $\sigma$ are 
 assumed to affect the log-price in a linear manner (via the change in
 demand)  modifying the
 discretized pricing formula \eqref{price2} as follows
  \begin{equation}
p(n)=p(n-1) \exp\left(\sqrt{h}\eta(n)
 -h/2+ \kappa\Delta\sigma(n)\right)\label{price3}
\end{equation}
where $\kappa > 0$ and   $\Delta \sigma (n) = \sigma (n) - \sigma(n-1)$.
The parameter $\kappa$ is a measure of the total market depth of the
slow agents.

At this point we note that \eqref{price2} can be recovered from
\eqref{price3} in two different ways. We can set $\kappa =0$
so that there are no slow agents, or we can suppose
that the slow agents are also, on average, always correct and $\sigma(n) =0 \;
\forall n$. In either case the models will generate the same 
price. However, the pricing formula \eqref{price3} allows for the
possibility of endogenous dynamics amongst the slow agents
affecting the price $p(n)$. Thus one can interpret
\eqref{price3} as stating that price changes have an exogenous
component $\sqrt{h}\eta(n) 
 -h/2$ due to new information, and an endogenous one,
 $\kappa\Delta\sigma(n)$,
caused by  internal complex dynamics.

Finally, we introduce one further
generalization of \eqref{price2} by weakening the assumption that the
fast (information driven) agents must always perfectly translate new
information into price changes. This is achieved by adding a (for now
unspecified) 
function $f(\bullet)$ that modifies the effect on prices of new information
entering the market via
  \begin{equation}
p(n)=p(n-1) \exp\left(\left(\sqrt{h}\eta(n) - h/2\right)
f(\bullet ) + \kappa \Delta\sigma(n) \right).\label{price4}
\end{equation}
with  the fast agents acting perfectly if $f \equiv 1$.

Before continuing, the differences between the fast and slow agents
should be clarified, especially since only the slow ones will be
simulated directly. In the numerics that follow, $h$ will be
chosen to  correspond to approximately 1/10 of a trading day.
Fast agents include institutions (or individuals)  that regularly trade the asset 
over a timescale of several days or
less, and/or are motivated primarily by new information. Slow traders
on the other hand will typically shift investment positions over weeks,
months or longer.

Equation \eqref{price4} does not yet constitute a closed system
because no rules governing the switching of the slow agents between
the $0$  and $+1$ states have been specified yet. There are many types of rule or
trading strategy that could be used, involving any desired combination
of pure utility function
maximization, bonus/commission maximization, inductive learning, imitation among a network of slow
agents, technical trading, `gut instinct', profits, losses,
relative performance of other investment options, market volatility,
fear, greed, margin calls, the weather, and so 
on ad nauseam. Prior studies such as the Santa Fe model (see \cite{lap99}) do indeed use
complicated  ecosystems of trading strategies and there is much to
commend this approach.

\begin{figure}[ht]
\vspace{0.2in}
\centering
\includegraphics[scale=0.5]{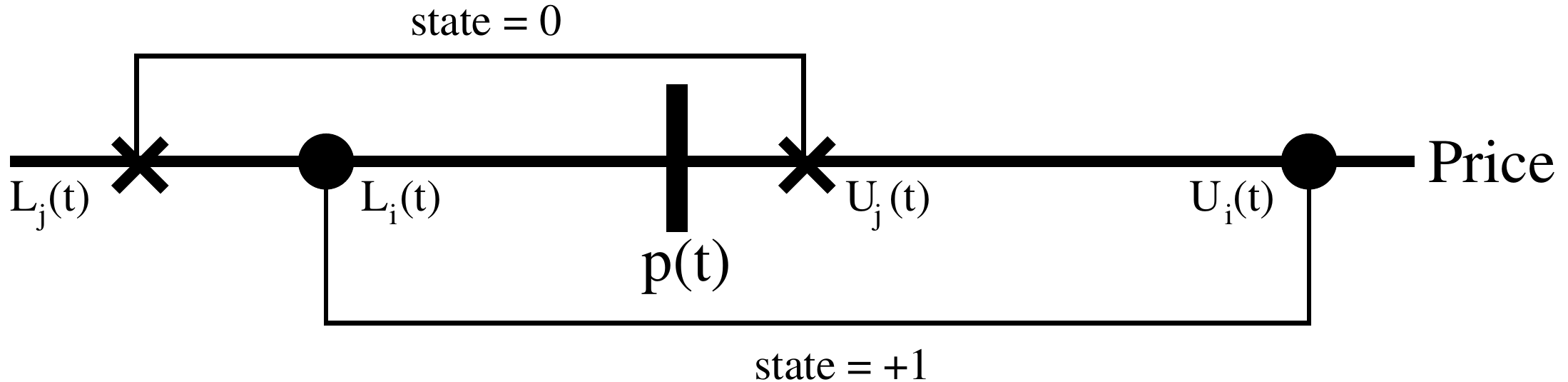}
  \caption{A representation of the model showing two agents in
    opposite states. Agent $i$ is in the $+1$ state and is represented by the
  (interval between) the two circles, and  agent $j$ is in the $0$
  state and is represented by the two crosses.}
  \protect{\label{fig1}}
\end{figure}

We shall use an approach based upon {\em moving price thresholds}
developed in \cite{ls08} and \cite{l10} which is deceptively simple
but capable of mimicking 
various real world influences, market `imperfections' and
psychological biases (see Section~\ref{mimic}). At the start of the $n^{\rm th}$ time interval, 
 the $i^{\rm th}$ slow agent is represented by its state, $0$ or $
1$, and an {\em evolving} closed price interval $I_i(n) = [L_i(n), U_i(n)]$ where $L_i(n) \leq p(t)
\leq U_i(t)$ (Figure~\ref{fig1}). The endpoints $L_i$ and $U_i$ are
referred to as the lower and upper thresholds respectively. If, at the end of that time
interval, the price  has  crossed {\em either}
threshold\footnote{For example, an agent who is $+1$ may be switching to
  either take profits or cut losses depending on which threshold is
  breached. An agent who is $0$ may be motivated either by a sell-off
  which makes them think the stock is now cheap, or an urge to follow
  the momentum of a rising stock or attempt to catch up with a benchmark.}, 
agent $i$ is deemed to be no longer comfortable with her
current investment position, switches states,  and the interval
$I_i$ is updated so that the price  is again an interior
point. Furthermore, from \eqref{price4} 
the action of switching causes a small jump in the log-price of
$2\kappa w_i/W$. Each of the $M$ slow agents has their own interval
straddling the current price 
with the price $p$ and all of the thresholds evolving at each
timestep.

Since the intervals $I_i$ are allowed to evolve, it just remains to define
the dynamics of the $L_i$ and $U_i$.
These  thresholds for each agent will change (usually slowly) between switchings
and correspond to that agent's
evolving  {\em strategy}. Or equivalently the intervals can be thought
of as their agents' comfort zones within which that agent is
still satisfied  with their current investment position.
Note that in the case of an
algorithmic `black box' or a hyper-rational utility maximizing investor these thresholds
will be consciously and explicitly known by that agent but no
other. A less-rational slow agent  may not even be consciously aware of the
threshold values but will know when one of them is violated and act
to switch. 

The state of the equilibrium model \eqref{price2} at any given moment
is completely specified by the current price (or equivalently the value of the
information stream $B(t)$). The situation for the full threshold model
is very different. To specify the current state of the system
completely requires the additional knowledge of all of the $2M$ threshold
values  and the rules specifying their dynamics. These endogenous `hidden
variables' add a great deal of potential complexity to the
model but we shall make some simplifying assumptions that appear to be
sufficient to address the stability issues at hand.

In order to apply an averaging argument and preserve geometric
Brownian motion pricing \eqref{price2}, we suppose 
that
there are myriad influences on slow agents' strategies that can  be adequately
represented by different (and uncorrelated, independent) geometric
Brownian motions applied directly to every agent's 
thresholds. These will be a mixture of purely rational independent analysis,
completely irrational thought processes, and mixtures of the two
almost certainly 
in differing proportions for each agent. Each of these influences will move
each upper and lower threshold either inwards or outwards, towards or
away from, the
current price thus making the agent
more or less likely to switch states respectively. Thus the threshold
dynamics between switchings are given by
\begin{equation} L_i(n+1) = L_i(n)  + p(n){\cal N}(0,h\delta_i), \quad 
 U_i(n+1) = U_i(n) + p(n){\cal N}(0,h\delta_i).\label{maj} \end{equation}
The quantity $\delta_i$ is the volatility of the threshold motion per
unit time for the $i^{\rm th}$ agent.
Finally, if an agent does in fact switch at the end of a time
interval,  their threshold values will reset
in such a way that is also independent of the other agents.
Thus, on average, equal numbers of slow agents will be
switching at each timestep and, provided that $\sigma(0)= 0 $ with identical
threshold distributions for the agents in each state at time 0,
$\sigma$ remains close to 0 for all time resulting in the standard equilibrium pricing
model \eqref{price2}.

We have now, finally, reached the point where we can endogenously perturb the
price. To do this we need to include
an influence on the slow agents that cannot be averaged away.
We posit that agents who are in the minority
state (e.g. those who are in state $+1$ if $\sigma < 0$) will, on
average, have a motivation to switch and join  the
majority. Furthermore the pressure to join the majority increases with
the magnitude of the difference between the two groups, measured by
the quantity $|\sigma|$. There are
several effects that naturally lead to such a perturbation. 

Firstly, any change in $\sigma$ will lead to a drift in price that may
be (mis)interpreted as a fundamental trend with the minority agents reacting
accordingly to the price signal.
Secondly, there may be `rational herding' by
investment managers who find themselves chasing a benchmark average
so as not to lose their jobs, bonuses or  investment capital
\footnote{These effects will be amplified by the short time-horizons of
such evaluation periods.}(see \cite{k36,ay85,ss90}). A third cause is 
 the actions of momentum investors who are consciously trying
to take detect a nascent bubble as part of their investment strategy.
Fourthly, there is the propensity of people, faced with uncertainty,
to believe that other people are better informed than they are,
 or their preference  to risk failing  conventionally than
succeeding unconventionally (\cite{k36}).
Finally, there may be purely psychological effects at work caused by
the discomfort of being in a minority, especially within social or
professional networks. We shall refer to all the above effects as
causes of {\em herding}.

For simplicity we assume that each agent reacts to the current
value of the aggregate (excess) demand/sentiment, $\sigma$, although in
reality  agents will have different perceived values of this quantity
(or may be 
 be reacting, in part, subconsciously). 
Herding is introduced into the model by supposing that for agents {\em in the
minority position only} \eqref{maj} is replaced by 
\begin{eqnarray}  L_i(n+1) &=& L_i(n) + p(n)(C_i h |\sigma(n)| + {\cal
    N}(0,h\delta_i)) \nonumber  \\
 U_i(n+1) &=&  U_i(n) - p(n)(C_i h |\sigma(n)| + {\cal
   N}(0,h\delta_i))\label{min} \end{eqnarray}
while those in the majority remain unaffected. Note that the change is
simply to suppose that each minority agent has an inward threshold
drift added to the dynamics. Drifting the thresholds inwards (towards
the current price of course) reduces the time to the next switching
and can be described as the agent's comfort zone being `squeezed' by
the majority opinion. The rate of drift is taken to be proportional to the length
of the timestep, $h$, the magnitude of the imbalance  $|\sigma(n)|$ and
a constant $C_i \geq 0$ quantifying the herding effect for that
agent. Note that herding is a positive-feedback effect --- as $\sigma$ moves
away from $0$ it, at least initially, provides a  mechanism for the
imbalance to increase. This completes the description of the model.

\subsection{Justifications for moving thresholds} \label{mimic}

As mentioned above, once equation \eqref{price4} has been reached one
can in principle use any modelling paradigm to specify the switching
rules of the slow agents. Treating each slow agent as a dynamic closed  interval
$[L_i, U_i]$ on the positive real line (that must at all times contain
the current price) is certainly unusual and at first may seem highly
unnatural. However the use of pairs of price thresholds offers some compelling
advantages.

Firstly there is the observation that slow agents react mostly to
price changes over longer time periods  rather than the arrival of 
new information. Indeed investment advice
and analysis is usually offered in the form of price triggers, as are
the outputs of computerized trading algorithms. Sometimes, such as in
a margin call, the agent has no choice over the pricing point.

A second important issue is that of transaction (or sunk)
costs. These are often neglected under simplifying assumptions but they
profoundly change the nature of agents' behavior\footnote{It is
  amusing to note that many of the people who rely on such models are
  actually paid from sunk costs. And all-too-often the possibility of
  significant 
  transaction fees can skew the information and research
  entering the market.}. Even if one
believes that agents are continuously maximizing their utility
functions this must somehow be translated into an acceptably small
number of trades since a  continuous process of incremental
adjustments would be ruinous. Provided that switching results in
new threshold values that are a non-zero distance away from the current
price, then moving price thresholds are a potential
mechanism for converting one into the other. The existence  of sunk costs
is closely linked to issues of hysteresis, memory-dependence and
non-reversibility that will be discussed further in Section~\ref{mem}.

Finally we turn to behavioral economics. It has already been shown
how the propensity for herding (rational and irrational) can be
included by moving thresholds inwards for agents in the minority
position. However other effects such as anchoring and loss-aversion
can also be replicated. Anchoring is almost automatic as thresholds
are reset around the last trading price while loss aversion requires
slightly more complex rules involving keeping track of whether agents
have made a profit or a loss. An extreme, but unfortunately quite
common, example demonstrates the idea.

Imagine an individual who bought  a dotcom stock at the height of
the tech bubble. Immediately the price goes down but, due to loss
aversion which is  the emotional difficulty of acting to realize a loss, the lower
threshold moves down even faster. It is likely that the upper threshold is
moving downwards too but there is never enough of a temporary bounce
in the stock price to  cause a switch. Eventually the stock price hits
zero without the agent ever selling.

Recent work (see \cite{kt74,r98,g02a,bt03,e07}) has suggested plausible heuristics that
agents may be using in practice. It should be possible to recreate such rules
using moving thresholds and then observe the effects upon aggregate statistics.

\section{\label{num1}Preliminary Numerics}

Detailed numerical
investigations from a similar model (using two pairs of
static thresholds for each agent) are compared against the stylized
facts in \cite{ls06} and further numerical results for a  moving 
threshold  model can be found in
\cite{ls08}. Here, for completeness,
we provide enough details for
replication of the numerical results and sufficient simulations to reveal the nature of
the non-equilibrium solution. It must be emphasized that no fine-tuning of parameters is
required.
Most of the parameters can be roughly estimated and we do so
conservatively and as simply as possible. 

Firstly, we must choose a timestep $h$ defined in units such that the  variance of the
external information stream is unity for $h=1$. 
An observed  daily variance in price returns of 0.6--0.7\%
suggests that $h=0.000004$ should correspond to approximately 1/10 of
a trading day. The price changes  of 10 consecutive timesteps are then summed to give
the daily price return. 

We next assume that all slow agents have equal weight $w_i=1$. This
could be replaced by a Pareto distribution but we shall not do so
here\footnote{It is not necessary to assume that each slow agent
  corresponds to just one individual or institution --- they could
  each refer to a subset of agents with very similar strategies or propensities.}. The
behavior of the system is largely independent of the number $M$ of
slow agents with $M=1000$ being a lower-bound for representative
simulations. All simulations in this paper will use $M=100000$ slow
agents.

Next we fix how the slow agents' thresholds are reset after a switching.    
If agent $i$ switches at a price $p^*$ then immediately afterwards
 the interval is reset to 
$$[L_i,U_i] = [p^*/(1+Z_L),
p^*(1+Z_U)]$$ where $Z_L,Z_U$ are each chosen from the uniform
distribution on the interval $ [0.05,0.25]$. This 
corresponds to an initial strategy that requires price ranges 
in the range 5--25\% before another switching (although of course the threshold
dynamics will alter the strategy as the system evolves).

We now turn to the most significant parameters, the herding parameters
$C_i$ for each agent. Let us consider the more herding-susceptible agents
and estimate that they can be pressured to switch over a period less
than a reporting quarter (about 80 trading days). This is certainly the case for fund
managers who are trying to keep up with a performance benchmark, or
momentum investors, say, who try to time the market several times a
year. Then a simple calculation based upon \eqref{min}
suggests $C_u=100$ as a reasonable  upper bound for the herding parameters. Thus
we assign agents with a herding parameter chose uniformly from the
uniform distribution on $[25,100]$ with the somewhat arbitrary but
unimportant lower
bound corresponding to agents who are relatively immune to herding
effects.  

The parameters $\delta_i$ represent the `volatility' of each agents
strategy (or expectations) and we simply assume they are all
equal. Note that if $\delta_i=1$ then the volatility of the thresholds
is the same as the volatility in price due to the information $\sqrt{h} \eta$
entering the 
system in \eqref{price4}. This is probably too large since the slow
agents are not 
the ones motivated by new information and should alter their
expectations more slowly. Thus we set $\delta_i = 0.2 \;
\forall i$.    

Finally, as regards the slow agents,
 we come to the parameter $\kappa$ that represents the effect
of aggregate slow demand upon the price in  \eqref{price4} 
but which is much harder to estimate a
priori. Simulations (not presented) show that even with $\kappa =0$
endogenous dynamics and loss of equilibrium exist but without
affecting the stock price at all. However if $\kappa$ is too large the effect
upon the price is so severe that the resulting price fluctuations are
unrealistically large. However for a surprisingly wide range,$\kappa
\in [0.1,0.3]$, the price statistics are in good general agreement
with the stylized facts. We conservatively choose $\kappa = 0.1$ as
our default value in the simulations.

We now consider the fast agents. The function $f(\bullet)$ was
introduced in   \eqref{price4} to modify the effects of the fast
agents by assuming that under certain circumstances they do not
accurately translate new information into price changes\footnote{This is a
simple but plausible mechanism for the generation of volatility
clustering.}.We shall suppose
that at times of extreme market sentiment, when  $\sigma$ is far away from
$0$, excess speculation by fast traders occurs. This may be due to new
agents entering the market or by too much attention being paid to new
information by
traders expecting a market correction. There is some evidence for this
(see \cite{b99}) and it also helps correct for the fact that in our simple model the slow
agents are not allowed to own  multiple units of stock. As in previous
work on this model the simplistic but plausible choice $f(\bullet) = 1
+ \alpha |\sigma|$ is made with $\alpha >0$. We choose $\alpha = 1$ so
that at the most extreme mispricings, new information moves the market
twice as much as it would if the fast agents were correctly
incorporating it. It must be stressed that the presence of the
function $f(\bullet)$ has no effect upon the main conclusions of this paper
and will be set to $f\equiv 1$ for one of the simulations in
Section~\ref{num2} to demonstrate this.

\begin{figure}[ht]
\centering
\includegraphics[scale=0.3]{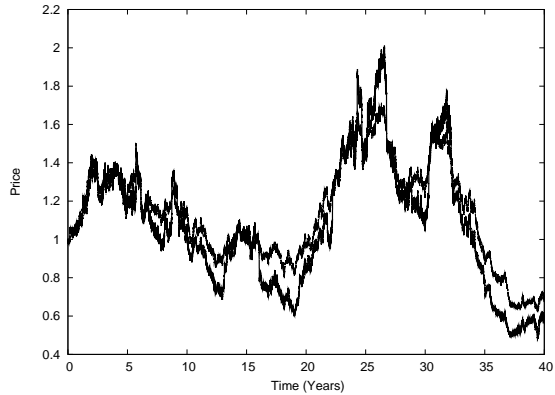}
  \caption{ Asset price of a simulation over 40 years with 100000 agents.}
  \protect{\label{fig2}}
\end{figure}

\begin{figure}
\centering
\includegraphics[scale=0.3]{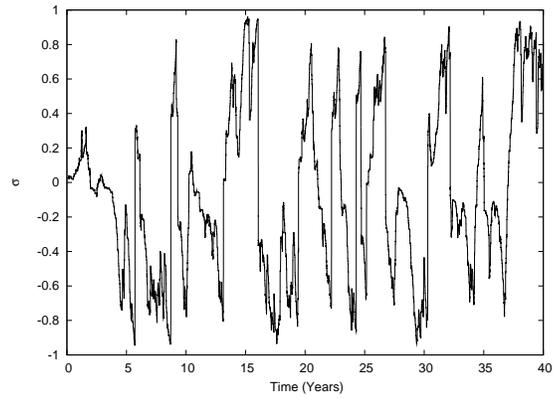}
  \caption{ A plot of $\sigma$ against time for the simulation in Figure 2.}
  \protect{\label{fig3}}
\end{figure}

 Figure~\ref{fig2} shows a simulation with the above parameters. The
 initial states of the agents and their thresholds are randomized. The
 thicker, more volatile curve is the pricing output of \eqref{price4}
 while the lighter curve is the output of the geometric Brownian
 pricing model (that, recall, can be recovered by either setting
 $\kappa = 0$ or all of the $C_i=0$). A typical characteristic is that
 mispricings develop 
 slowly and then suddenly reverse. This can also seen in Figure~\ref{fig3}
which plots $\sigma$ against time.
 
\begin{figure}
\centering
\includegraphics[scale=0.3]{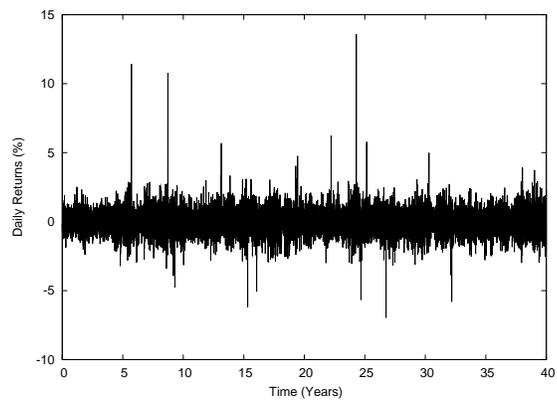}
  \caption{Daily price returns for Figure 2.}
  \protect{\label{fig4}}
\end{figure}

\begin{figure}
\centering
\includegraphics[scale=0.3]{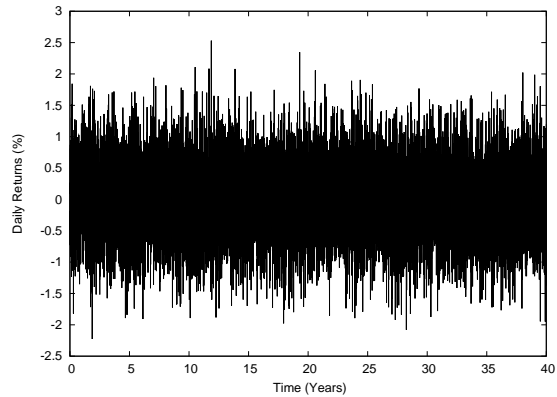}
  \caption{Daily price returns of the equilibrium price from Figure 2.}
  \protect{\label{fig4a}}
\end{figure}

Both parts of  the mispricing sequence are caused by   {\em endogenous}
dynamics.
The incremental mispricing is due to herding effects. The sudden reversals
occur because eventually enough of the majority agents switch position
to start a cascade process --- as agents switch they cause a change in
price (due to the $\kappa \Delta \sigma$ term in \eqref{price4}) that
trips other agents' thresholds and so on. 
Figure~\ref{fig4} shows
the daily percentage price returns. There are a significant number of
large price changes that cannot be explained by a Gaussian information
stream and lie within the fat tails\footnote{Interestingly there is a
very close correspondence between the dynamics of these cascades and
Queueing Theory that is described in detail  in \cite{l10}.}. For
comparison, the price changes from the equilibrium model are
shown in Figure~\ref{fig4a}.

\begin{figure}
\centering
\includegraphics[scale=0.3]{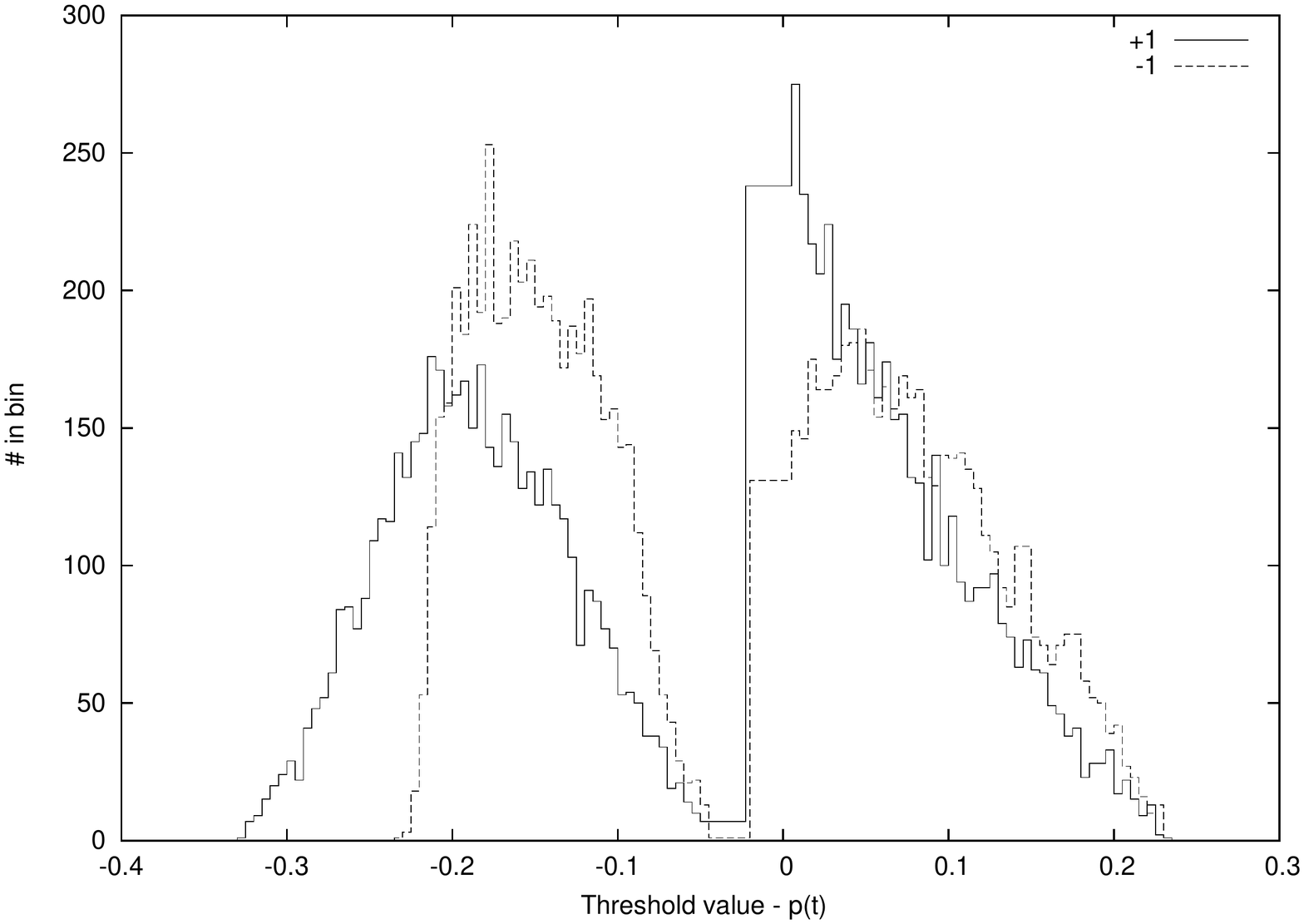}
  \caption{ The density of the thresholds along the price axis
 for agents in each of the
    two states. The
    difference between 
    the two distributions is a `memory effect' of the prior behavior
    of the system and affects the future evolution.}
  \protect{\label{fig5}}
\end{figure}

A snapshot of the internal structure of the market is shown in
Figure~\ref{fig5}. The density of the lower and upper thresholds of
each type of agent ($0$ and $+1$) are plotted relative to the current
price at a moment when $\sigma \approx 0$ (the mismatch between the
densities is far more severe when $|\sigma| \approx 1$). The two density plots are
not identical (as they were in the initial state) meaning that as the
system evolves $\sigma$ will once again move away from $0$ because
different numbers of agents will be switching in either direction.
The question that will be asked in Section~\ref{num2} is: how strong
must the herding effect be to generate significant endogenous dynamics?  

\section{\label{links}Links to economic concepts and extant  
models}

\subsection{Equilibria, memory and history-dependence}\label{mem}

The notion of equilibrium described in the introduction is a very strong
one, precluding the possibility of multiple  internal configurations
for the same external parameters. Nonetheless the absence of
(non-trivial) endogenous dynamics is a prerequisite for models that
are {\em reversible} and history-independent with temporary exogenous
shocks having no permanent effect.  

The key issue is one of {\em memory} at the micro-level. If individual
agents have no memory then it becomes much easier to assume that they
will reverse their actions and expectations when the external
influence is removed. However
economic actors are subject to many effects (both rational and
irrational) 
that cannot reasonably be
modelled in this way.

Perhaps the most significant of these rational factors is the ubiquitous presence of
transaction (sunk) costs (see \cite{d92,pglc99,g02}). These are
expenses incurred that cannot be
recouped on reversing the action. 
As an example suppose that at
the current widget price it is not profitable for a
manufacturer to have a factory produce the widget. However when the price
increases to $\beta$ (perhaps due to a demand shock)
  the firm switches a factory over to widget 
production from something else, incurring costs such as re-tooling and
factory down-time. If the price then falls back below $\beta$
the factory will not 
immediately switch out of production but rather waits until the price falls
below some value $\alpha$. Thus if one only looks at the
current price $p$ and $\alpha < p < \beta$ it is not possible to know
what the factory is producing --- one also needs to know which of the
threshold values $\alpha$ and $\beta$ was last crossed.

In the physical sciences this is referred to as {\em
  hysteresis}\footnote{In economics the term `hysteresis' is often
  taken to mean the persistence of deviations from equilibrium.} and
the reader is directed to \cite{cgl09} for a fuller 
description of the role of hysteresis in economics. 
The presence of many such factories, all with differing threshold
values, results in many possible alternative internal configurations for
the same price level. Each of these  possibilities results in
a different future
evolution of the system which now displays both irreversibility and
history-dependence\footnote{When such effects are observed in
  macroeconomics a common  equilibrium-based 
explanation is the presence of a  {\em unit-root} since, if a system is only
marginally stable  it will take a long time to return to its
former value after a disturbance. There are standard tests, under
assumptions of linearity and the absence of hysteresis, for
determining if this might be the case (see \cite{sd84}). However marginal
stability implies that a system is close to instability and this
should be far more worrying to economists than irreversibility.}.  

At an abstract level, the thresholds used to describe the slow agents
in Section~\ref{model}
are a mechanism for introducing memory/history into the modelling
process (the fast agents, by assumption, act upon new information and
require no such mechanism). 
As an example, an agent who has been in the minority state for a long
time and is influenced by a herding pressure, on average will have
thresholds that are much closer together and be more 
likely to switch in the near future.
This information is propagated from one timestep to
the next along with the agent's current state.

It is now worth revisiting the
concept of equilibrium, allowing for the possibility of multiple endogenous
configurations (and dynamics). 
Saying that the market model simulation from Section~\ref{num1} is
in equilibrium until just before a
sharp reversal is akin to saying that a geological fault-line is in
equilibrium until just before the earthquake. This is true,
in that there is a balance of both external and internal forces, 
but the statement that the fault-line is in equilibrium just after the
earthquake is equally true. The `before' and the `after' can be
approximated deceptively well by unique equilibrium models, but not
the transition!
 
The `balance of forces' notion of equilibrium is not
sufficient to guarantee uniqueness and the system can rapidly move from one   
internal state to a different one (with lower energy in the case of an
earthquake). Finally, it should be noted that multiple equilibrium models do
exist in mainstream economics with the initial conditions determining
which equilibrium is achieved. However the situation here is far
worse ---  the set of feasible final states cannot be enumerated in
advance and depend on the path taken by the process.

\subsection{Rational expectations and efficient markets}

The philosophical, political, social and practical consequences  of
the mainstream acceptance of the
hypothesis of memory-free, efficient, financial markets cannot be overstated. Although the
concepts were introduced by Bachelier in his 1900 Ph.D. thesis, this
work was  largely forgotten
until the 1960s when they became known collectively as the
Efficient Market  Hypothesis (EMH) (see \cite{f65,s65,f70}).

The three versions of the EMH (weak, semi-strong and strong) all rely upon
two qualitatively different classes of assumption. 
Firstly, there are strong assumptions about the market itself and the nature
of the information stream entering it. Such information consists of,
amongst other things, economic
statistics, performance 
reports, geopolitical events, and analysts’ projections. It is assumed
to be  instantly available
to all economic participants, uncorrelated with itself, and is usually
modeled as a Brownian 
motion, possibly with drift. 

The second class of assumptions relates to
the  market participants
themselves, who are deemed to be perfectly rational, correctly incentivized, and capable of
instantaneously incorporating
new data into their differing market strategies and
predictions. However, heterogeneity of agents (or their expectations) is 
necessary to ensure that trading occurs in the absence of arbitrage
opportunities.
Thus the final ingredient in the EMH description is the Rational Expectations
Hypothesis (REH) (see \cite{m61}) stating that the differing
expectations driving trades, when used 
as predictions, are on average 
correct\footnote{As opposed to being consistent with the modelling assumptions
which is the form of the REH most often used in macroeconomic
modelling.} and do not result in market mispricing. Additional
assumptions, such as the absence of 
transaction costs, yield the standard formulae used for risk
management and derivative pricing
which form the bedrock of modern financial engineering (\cite{bs73}).
It is this second class of assumptions that are the focus of this
paper.

As described  in Section~\ref{model}, if one sets $f(\bullet)\equiv 1$,
$C_i = 0\; \forall i$ and further assumes that all threshold dynamics
for the slow agents are the
result of perfectly rational, independent, utility maximizing behavior (satisfying the
REH) then one recovers an equilibrium market following  geometric
Brownian motion and satisfying the EMH. Once one weakens these
assumptions to allow for more 
general dynamics (and motivations), the moving threshold model can be
thought of as a `perturbation space' within which one can explore the
robustness/stability of the default EMH model.

Even if agents are not perfectly rational and other factors
influence the threshold dynamics, the pricing should remain correct provided
that the REH still holds. However the presence of just one REH-violating
perturbation  potentially invalidates its use. As was shown in the
numerical simulation of Section~\ref{num1}, real-world effects that induce
herding do exactly that, giving rise to price dynamics that
differ greatly (qualitatively and quantitatively) from  the
equilibrium model by introducing a form of coupling between agents' actions. However
it is possible that lower  
levels of herding may  result in acceptably small deviations from equilibrium. 
In Section~\ref{num2} the herding parameter will be systematically
varied to show that this does appear to be the case, although only if the
herding parameter is reduced by a significant factor from its
estimated real-world value.

Even markets in which irrationality
exists can be `efficient' in the sense that investors cannot earn
above-average returns without taking on above-average  risk.
Indeed, this is a minimal requirement for any predictive market
model (see \cite{m03})\footnote{The aim of this work is not to show that
irrational   markets are inefficient, rather that equilibrium models
are fundamentally inadequate.}. In \cite{cgls07} it was shown that there is no statistically
significant different between the the investment performance of agents
with differing herding propensities $C_i$ (and hence nothing to be
gained by adaptively changing their reaction to herding). Indeed when
transaction costs are taken into account the traders with the highest
values of $C_i$, that includes  momentum traders, performed
significantly worse, in agreement with previous studies (\cite{o99}).   

One potential criticism of the model is that the fast agents are
assumed to be reacting to {\em new} information and converting it into
price {\em changes}. However there may also exist true fundamental
fast agents who
are aware of the current non-zero value of $\sigma$ and the correct
price and who view this as an arbitrage opportunity. This would act as
an additional equilibrating (negative-feedback) mechanism helping to
counteract the herding. This brings us to the very important issue of
the limits of arbitrage, both in the model and in real markets.

Firstly, as regards the model, $\sigma$ is assumed to be precisely known by
all the slow agents. This is purely for simplicity and agents may have
widely-differing perceived values that are only approximately (or on
average) correct. Also, it is important to note that no agents are
assumed to know the correct gemetric Brownian motion price. It is
calculated and plotted in Section~\ref{num1} but this is only a visual
aid --- none of the agents need to be made aware of it. As it
stands the model is a caricature, albeit one that can be made
arbitrarily more complicated and realistic. As this complexity grows, any potential
model-specific opportunities for arbitrage that might exist will
reduce and so now we 
discuss the limits to arbitrage in  real markets.

Arbitrageurs and/or fundamentalist investors provide a possible
to counteract herding effects. However there are
severe limitations in practice. Firstly there is the noise-trader
problem (see \cite{s00} and \cite{sv97}) --- arbitrageurs typically have very short
time-horizons and mispricings can last a very long time. Secondly
there is the existence of speculative traders and short term
momentum-traders who may
actually make the mispricing worse. Thirdly, it is difficult in
pactice to be sure what the fundamental price actually is. There is no visible,
unambiguous, information stream and all trends may be misinterpreted as
rational --- especially by those who subscribe to the EMH!
Some evidence for this may
be found in the wide variations over time of even the most basic measures of
value such as the P/E ratio of a stock.   

This is not to dismiss entirely the possibility that herding effects
can be counteracted. Indeed many, perhaps even the vast majority
of, potential bubbles may get
deflated before anyone even noticed by rational agents working as the EMH
suggests they should. The point is that it does not happen every
time. The stability results to be presented in Section~\ref{num2} should, in this light,
be viewed as a preliminary attempt to quantify which effect eventually
wins. 

%bw07

%a posteriori, the sigma problem,

%information cascade expected utility , asymmetric payoff
%--- take cues from each other

%Theoretical work on information cascades by writers such as Diamond and
%Dybvig (1983), Bannerjee (1992) and Bikhchandani, Hirshleifer, and
%Welch (1992) 

%keynes general theory(scharfstein p1)
%In Scharfstein avoid feedback. receive signals, conditions for herding 
%unique equilibria, 

%shiller and pound (see schafstein).

%information cascade p190. rationally irrational. prices no longer
%reflect fundamentals.

%speculators riding a wave, make f(.) even more volatile, even more of
%a nonsense aka EMH. Noise-trading, p182. Momentum traders

%rational arbitrage can destabilize security prices

\subsection{Technical analysis}

Given the widespread belief in the underlying notion of (at least
weakly) efficient markets, a surprisingly large number of people are
employed in technical analysis, looking for exploitable trends and
patterns in past pricing/volume data. Some studies, such as \cite{bll92}, claim that the
most popular trading rules (based upon moving averages and
support/resistance levels) can indeed produce statistically
significant profits, even in the presence of transaction costs, while
others dispute this (see \cite{chan96} and \cite{m03}).

An obvious question to ask is what effect such technical traders might
have
on the market, if any. A second question is, if a particular 
technical trading rule does
indeed work,  what are the reasons for it? It may be caused by the
presence of one or more systemic defects (such as herding effects,
perverse incentives, or behavioral effects) or it may in fact be a
self-fulfilling prophesy caused by the large numbers of technical
traders using that rule themselves. By incorporating such strategies
into the threshold dynamics, it is possible to systematically
explore such questions. Preliminary results will be presented elsewhere.

\subsection{Soros' Theory of Reflexivity}

In \cite{soros03,soros08}, George Soros introduced his Theory of Reflexivity. While this
is still at the stage of being a
philosophical theory, and as yet has had little impact on economics
or finance,  its relationship to the modelling approach used
here (and between its predictions and the numerical results presented
above) is close enough to merit comment. 

Soros' theory rests upon two observations. Firstly, human beings are
fallible and they may misinterpret an apparent trend, or some
fundamental misconception may take hold. This results in investor
behavior that is incorrect but in turn induces changes to the market or economic
system. Positive feedback effects then  cause an increasing mismatch
between perception/prices and economic fundamentals that
eventually becomes unsustainable and rapidly unwinds. 

In the model of Section~\ref{model}, herding provides the 
amplification mechanism. This is reflexive in the sense that beliefs
affect prices, provided  $\kappa >0$ in \eqref{price4}.
Soros suggests that such processes are commonplace in economic
systems and result in far-from-equilibrium dynamics that only become
apparent at the very end. This is exactly what is observed in the
simulations from Section~\ref{num1}.

\subsection{Minsky and the financial instability hypothesis}

Recent events in financial markets have re-awakened interest in the
work of Hyman Minsky and in particular his theory of
financial instability (\cite{m74,m08b}). 
Minsky's work is unusual in macroeconomics in that it places a great
deal of importance upon the role of the financial system and debt accumulation. 
He also stressed, following on from \cite{k21},  the role played
by market sentiment (analogous to the quantity $\sigma$ above), belief under
uncertainty, systemic risks and contagion.

Minsky argued that a rising trend in prices (for whatever cause)
during a period of relative stability will attract savings and profits
leading to further price rises. Gradually there is a reduction in the 
perceived level of risk that encourages more lending, debt and
leverage. Lending standards fall, risk-taking increases, 
 and `Ponzi borrowers' appear who
are relying upon increasing prices to service their debt. Eventually the
system becomes unstable, credit tightens,  and prices cascade
downwards in a `Minsky moment'.   

Minsky believed that such processes were the norm rather than the
exception with disequilibrium adjustment mechanisms being insufficient
to counteract or prevent them. The resulting dynamics are 
 similar to those observed in the numerics of
 Section~\ref{num1} --- there is a long period of apparent trend
 stability at the aggregate level (whose magnitude and duration are
 extremely hard to predict)   that masks increasing {\em endogenous} instability.  
Then when the process ends it does so very rapidly. Or to put it
another way, similar positions develop gradually and then unwind quickly.

There are two ways in which the model presented here relates to the
work of Minsky. Firstly, the increasing availability of low-quality credit and
lowered perception of risk
provide yet another herding mechanism that adds `fuel to the
fire'. Secondly, if one reinterprets the price $p(n)$ in the model of
Section~\ref{model} as being a quantity that represents the overall
level of, and ease of obtaining, credit (with slow
agents being potential lenders) then one has a model that is
distinctly Minskian. A more sophisticated model might couple together
two such models, one for price and one for credit\footnote{It should
  be pointed out the model in this paper is symmetric with respect to
  rising and falling prices, while Minsky's arguments are not. This is
  not a fundamental problem, however.}.

%\subsection{Queueing Theory}

\section{\label{num2}Stability simulations --- how much herding is too
much?}

In the numerics of Section~\ref{num1} the herding parameters $C_i$ for the
slow agents were chosen from the uniform distribution on
$[25,100]$. However we know that if all the $C_i$ are set to $0$ then the
equilibrium solution is recovered. We now investigate the manner in
which the equilibrium solution loses stability as the herding is
systematically increased from 0.

We introduce a {\em bifurcation parameter} $C_{\rm max}$ and choose
all the $C_i$ uniformly from the interval $[C_{\rm max}/4,C_{\rm
  max}]$. In order to quantify the level of disequilibrium in the
system we record the maximum value of $|\sigma|$ over the last 30
years of the simulation (to remove any possible transient effects caused
by the initial conditions). 

\begin{figure}
\centering
\includegraphics[scale=0.6]{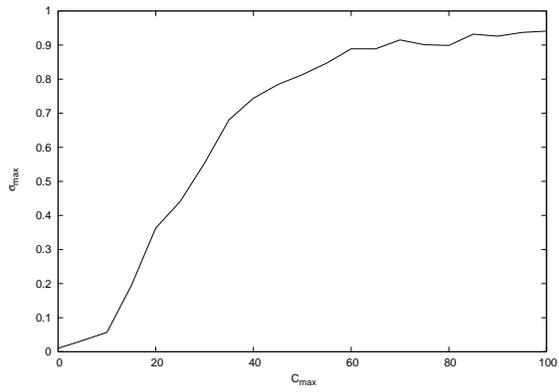}
  \caption{A plot showing the degree of disequilibrium $|\sigma|_{\rm max}$
    averaged over 10 runs for varying
    levels of the herding parameter $C_{\rm max}$.}
  \protect{\label{tab1}}
\end{figure}

Figure~\ref{tab1} shows the results (with the maximum of $|\sigma|$ averaged over 10
runs) with all other parameters kept unchanged 
from those used to generate Figure~\ref{fig1} --- the only difference
is that the initial value of $\sigma$ is set to 0.05 to provide an
initial disturbance to the system. 

As can be seen, the equilibrium solution can be considered  a good
approximation only
for $C_{\rm max} < 10$ which is an order of magnitude below our
(rather conservative) estimate $C_{\rm
  max}= 100$ from Section~\ref{num1}. The loss of stability, measured 
in terms of the maximum deviation of $\sigma$ from 0, is gradual and
saturates at around $C_{\rm max}= 60$.

A rough description of the dynamics of the system is as follows. 
The drift in the threshold dynamics of the minority agents is a
destabilizing influence, while the diffusion of the thresholds and the
fact that the majority agents do of course eventually switch out of
their position are stabilizing influences. The existence of such competing
forces is a very common cause of non-trivial dynamics in complex
nonlinear systems. A mathematical treatment of these stability issues
will be presented elsewhere.

\begin{figure}
\centering
\includegraphics[scale=0.6]{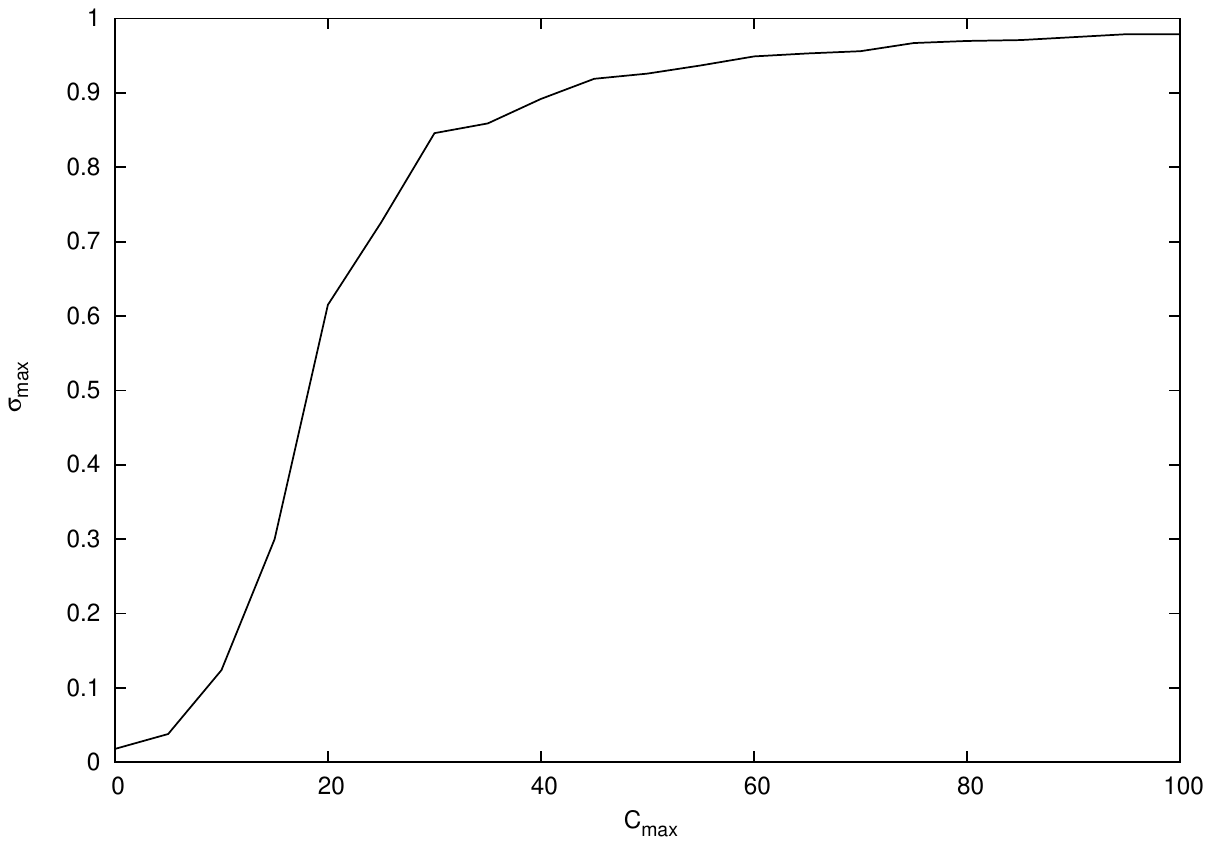}
  \caption{The same plot as \ref{tab1} but with $\alpha = 0$ implying
    that the fast agents are perfectly translating new information
    into price changes.}
  \protect{\label{tab2}}
\end{figure}

It could be argued that the presence of $\alpha \neq 0$ modifies
the influence of the fast agents and may be in part responsible for
the loss of equilibrium stability. This is ruled out by 
Figure~\ref{tab2} which shows the results with $\alpha = 0$. If
anything the instability is more pronounced with the full extent of
far-from-equilibrium dynamics being reached at $C_{\rm max} \approx
40$. 

\begin{figure}
\centering
\includegraphics[scale=0.6]{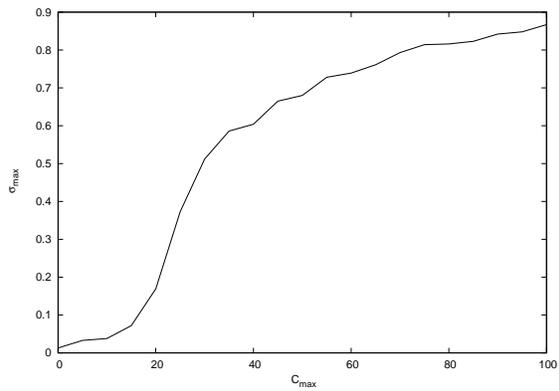}
  \caption{The threshold dynamics are made more volatile by increasing
    $\delta$ to 1.}
  \protect{\label{tab3}}
\end{figure}

Finally, we consider the parameters $\delta_i$ that govern the
diffusion of slow agent thresholds. In Section~\ref{num1} it was
argued that this would likely be lower than the price volatility and
so was set to $\delta_i = 0.2 \; \forall i$. Figure~\ref{tab3} 
sets $\delta_i = 1 \; \forall i$ so that the magnitudes are now
comparable. The change from Figure~\ref{tab1} is negligible.

\section{\label{conc}Conclusions}

The numerical results of Section~\ref{num2} demonstrate that a
hypothetical, yet recognizable, equilibrium  market model loses
stability in the presence of even relatively small herding pressures
--- whatever their underlying cause may be. Or to put it another way,
the positive feedback mechanisms caused by herding effects can
overwhelm self-correcting equilibrium models. This does not by
itself prove that such an instability occurs  in any
actual  market, but there are two observations that suggest this is
the  case. Firstly, the price statistics of the unstable
system  are much  closer to the stylized facts of financial markets
than those of the 
equilibrium model. Secondly, while the model itself is quantitative
and new, it shares
features with qualitative and long-standing critiques of equilibrium models and
neoclassical economics. Hopefully the work presented here will make a useful
contribution to the debate.

Suppose for a moment that herding is indeed responsible for
moving markets away from equilibrium. This has important
policy implications (apart from the desirable reduction in overall leverage that
is an obvious remedy for underestimating the risk from
fat-tail events). The global financial crisis that finally became apparent
to everyone  in
2007/8 appears to have been, at least in part, caused by the increased
attention to short timescales caused by, for example,  fund-managers chasing the
average every quarter, individuals trying to flip houses, and inappropriate short-term
performance-related bonuses  and 
upfront commissions being paid out throughout
the financial system\footnote{This general question of how complex
  systems with no natural timescale react to forces with an artificial
  time horizon may also be useful in other fields such as ecology.}. 
Regulations to increase the time horizon of financial actors would
reduce the herding pressure and indeed at the time of writing such
proposals are being either discussed or enacted in the US and the UK.

The mathematical description of the threshold dynamics lies within  a new
class of stochastic partial differential equations for which
closed-form analytic solutions almost certainly do not exist. This may
be unacceptable to many economists. However, the value of such models
(and solutions) in almost
every other
quantitative discipline is beyond dispute and there is no compelling
reason to suppose that economic systems should be an
exception.
For example, chaos theory has shown
that even very simple dynamical systems, evolving without any external influences, 
can be  inherently unpredictable 
(except possibly in a statistical sense). Yet, even in the absence of
explicit  solutions, chaos theory is still capable of quantifying both
the probabilities of events and the theoretical time limit on
meaningful predictions.

Finally, it is worth stepping back  to look abstractly at the
process leading from \eqref{price2} to the full moving threshold model.
This started with the transition from an equilibrium model 
 that implicitly relies upon averaging and a 
representative agent to a bigger system that explicitly allows for the
possibility of endogenous dynamics. But, if the rules governing the agents' dynamics are
assumed to be uncorrelated and independent then the aggregate
behavior of the system is 
unchanged. Then more  complicated dynamics 
for individual agents can be  introduced that correspond to the effect
under investigation and can be regarded as perturbations to the
equilibrium model\footnote{The use of thresholds to specify agent
  dynamics seems a very natural choice but is not required. Other
  types of rule may have their advantages.}.
Dynamic Stochastic General Equilibrium (DSGE) models may provide an
interesting starting point for a similar procedure as they also rely heavily
upon the use of averaging, representative agents and equilibrium
theory. Furthermore, the new-Keynesian DSGE models incorporate the
concept of `sticky prices' which suggests that the use of thresholds
may again be a suitable way to introduce non-standard effects within them.

\bibliographystyle{aer}
\bibliography{econ}

\end{document}